\documentclass[twocolumn,runningheads]{svjour2}
\smartqed  % flush right qed marks, e.g. at end of proof
\usepackage{graphicx}
\usepackage{mathptmx}      % use Times fonts if available on your TeX system
\journalname{Astrophysics and Space Science}
\begin{document}

\title{Gamma-ray binaries}



\author{I. F. Mirabel}

\authorrunning{I. F. Mirabel} % if too long for running head

\institute{I. F. Mirabel \at
              European Southern Observatory (on leave from CEA, France).
              Alonso de C\'ordova 3107,
              Santiago 19, Chile\\
              Tel.: +562 4633260\\
              Fax: +562 2075046\\
              \email{fmirabel@eso.org}           %  \\
}

\date{Received: date / Accepted: date}

\maketitle

\begin{abstract}

Recent observations have shown that some compact stellar binaries radiate the
highest energy light in the universe. The challenge has been to determine the
nature of the compact object and whether the very high energy gamma-rays are
ultimately powered by pulsar winds or relativistic jets. Multiwavelength
observations have shown that one of the three gamma-ray binaries known so far,
PSR~B1259$-$63, is a neutron star binary and that the very energetic gamma-rays
from this source and from another gamma-ray binary, LS~I~+61~303, may be
produced by the interaction of pulsar winds with the wind from the companion
star. At this time it is an open question whether the third gamma-ray binary,
LS~5039, is also powered by a pulsar wind or a microquasar jet, where
relativistic particles in collimated jets would boost the energy of the wind
from the stellar companion to TeV energies.

%\PACS{First \and Second \and More}
\end{abstract}

\begin{figure*}
\centering
\includegraphics[width=0.5\textwidth]{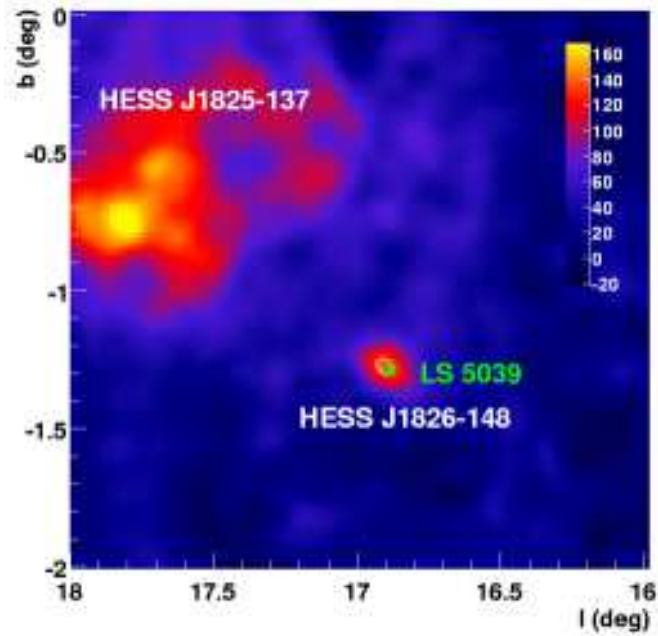}
\caption{Gamma-ray emission from the compact binary LS~5030 (from \cite{aharonian05a}). The green star indicates the position of the compact binary.}
\label{fig:ls5039} % Give a unique label
\end{figure*}

\begin{figure*}
\centering
\includegraphics[width=1.0\textwidth]{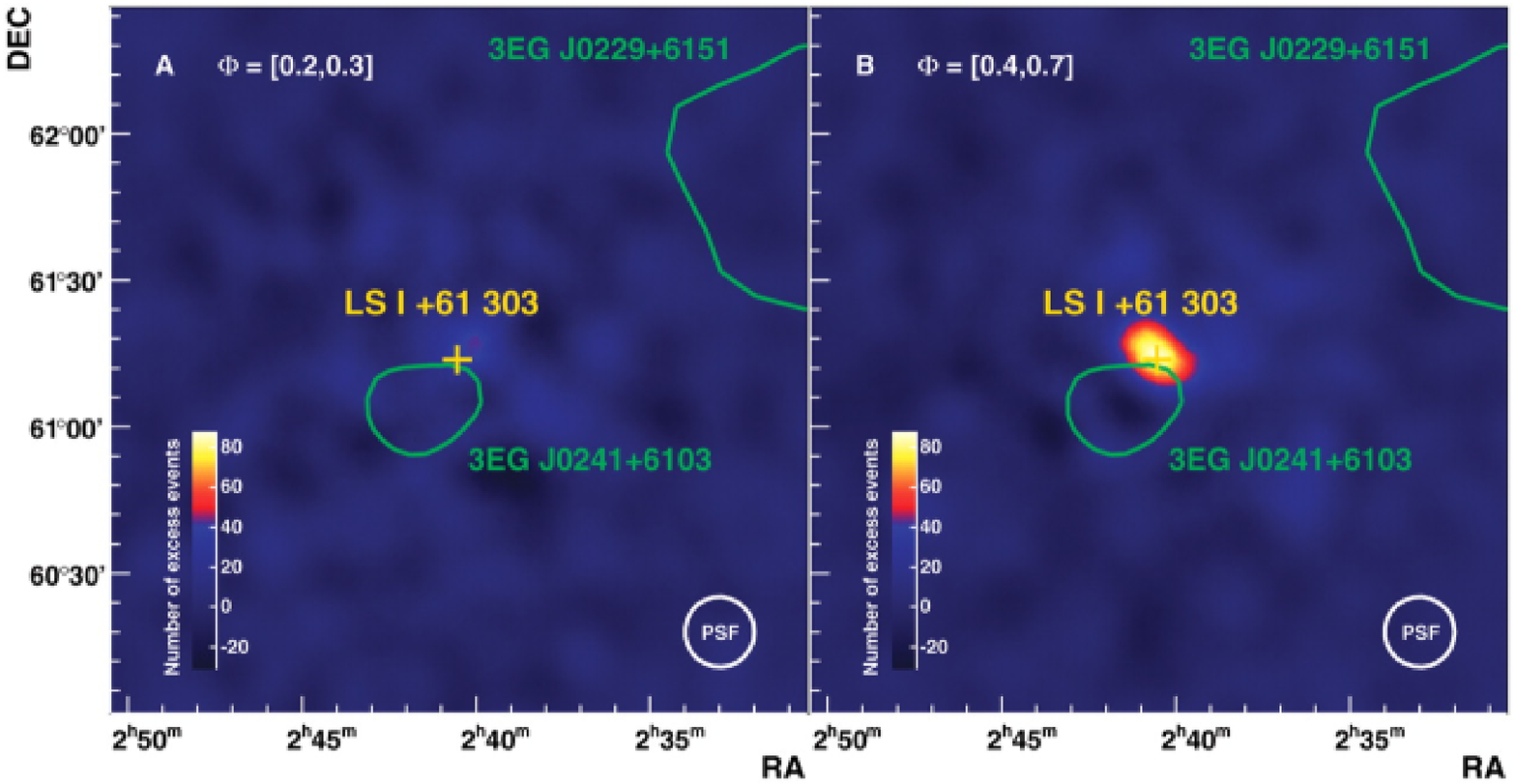}
\caption{Variable gamma-ray emission from the compact binary LS~I~+61~303 (from
\cite{albert06}). The crosses indicate the position of the compact binary.}
\label{fig:lsi} % Give a unique label
\end{figure*}

A new window on the universe is presently being opened by ground-based
telescopes that survey the sky by detecting very high energy (VHE) photons,
which have energies greater than 100 gigaelectron volts (GeV). Because of their
high sensitivity, and high angular and energy resolution, these telescopes are
revealing and identifying a plethora of new extragalactic and galactic sources
of VHE radiation. The Galactic Center, supernovae remnants, pulsar-wind
nebulae, and a new class of binary stars called gamma-ray binaries have all
been identified as VHE sources in the Milky Way. LS~5039 \cite{paredes00} is a
new gamma-ray binary detected at VHE \cite{aharonian05a} (see
Fig.~\ref{fig:ls5039}). Recently, Albert et al. \cite{albert06} confirmed the
identification \cite{kniffen97} of LS~I~+61~303 as the third gamma-ray stellar
binary, reporting a time variability in the signal that points to the mechanism
for the VHE emission (see Fig.~\ref{fig:lsi}).

\begin{figure*}
\centering
\includegraphics[width=1.0\textwidth]{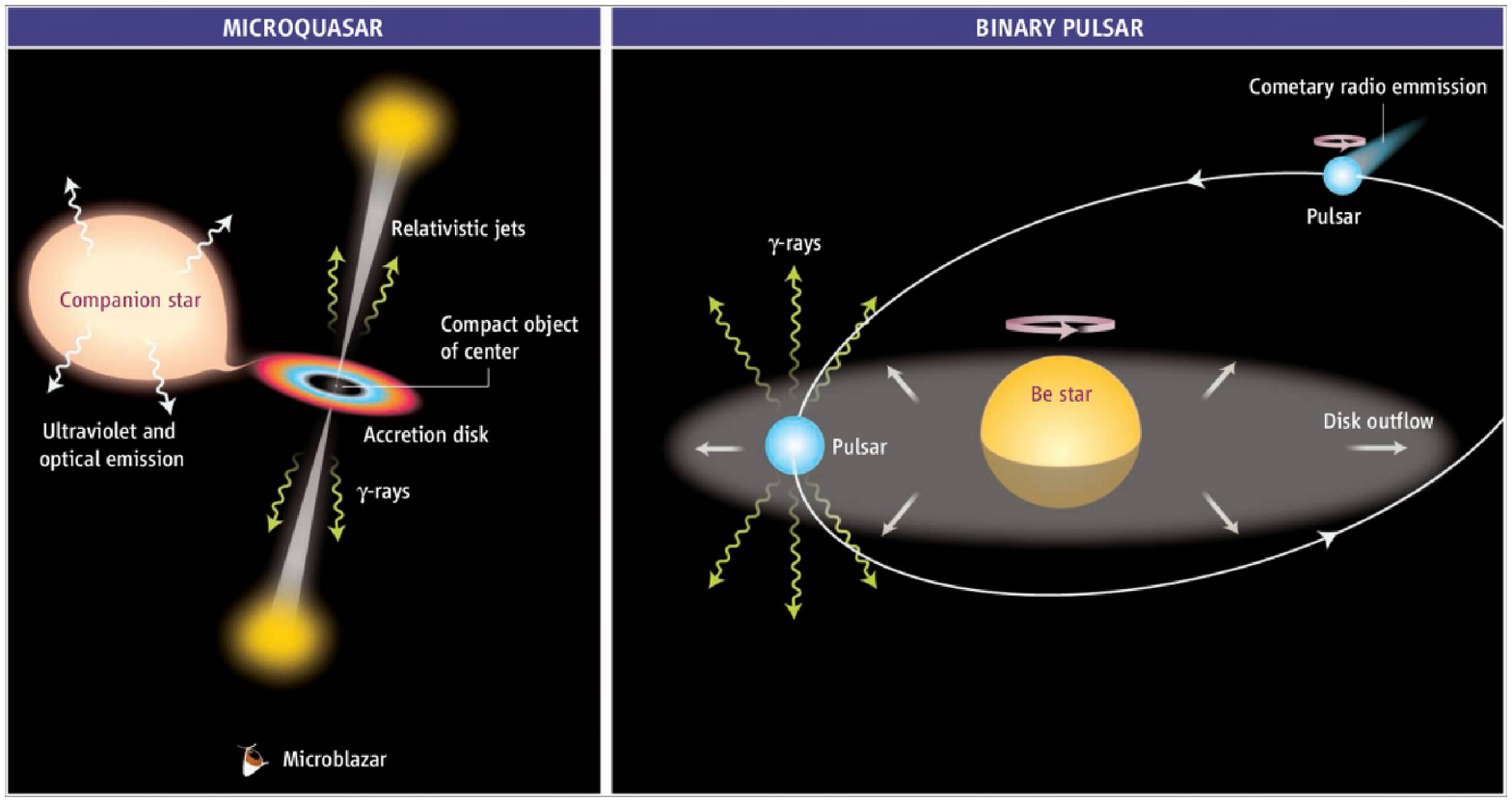}
\caption{Alternative models for very energetic gamma-ray binaries.
{\it Left}: Microquasars are powered by compact objects (neutron stars or
stellar-mass black holes) via mass accretion from a companion star. The jets
boost the energy of stellar winds to the range of very energetic gamma-rays. 
{\it Right}: Pulsar winds are powered by rotation of neutron stars; the wind
flows away to large distances in a comet-shape tail, as has been shown in
\cite{dhawan06} to be the case for LS~I~+61~303. Interaction of this wind with the companion-star outflow may produce very energetic gamma-rays.}
\label{fig:scenario} % Give a unique label
\end{figure*}

A microquasar-jet \cite{mirabel98} model (see Fig.~\ref{fig:scenario}, left
panel) has been proposed to account for the VHE emission from LS~5039. For
LS~I~+61~303, Albert et al. \cite{albert06} favored a mechanism, called inverse
Compton scattering, by which relativistic particles collide with stellar and/or
synchrotron photons and boost their energies to the VHE range
\cite{atoyan99,bosch06}. In the context of the microquasar jet hypothesis, an
alternative hadronic model has been proposed for the production of very
energetic gamma-rays \cite{romero05}. In this model the compact object accretes
matter from the dense and slow equatorial wind of a Be primary star. Gamma-ray
emission is originated from pp interactions between relativistic protons in the
jet and cold protons from the stellar wind.

Microquasar jet models for the production of VHE photons were a natural
hypothesis because VHE emission is also being detected from blazars, namely,
active galactic nuclei (AGN) whose jets are closely aligned with our line of
sight. Because the particle energy in microquasar jets is comparable to that of
particles in AGN jets \cite{mirabel98}, it was expected that microquasars could
also produce very high energy gamma-rays \cite{mirabel99}. This idea had been
strengthened by observations showing that the kinetic power of microquasar jets
may be larger than 10$^{39}$~erg~s$^{-1}$, which is larger than the radiated
power \cite{gallo05}. Furthermore, microquasar jets trigger shocks where
electrons are accelerated up to TeV energies \cite{corbel02}, providing the
necessary conditions for VHE emission. However, it is believed that TeV photons
from blazars are produced by relativistic Doppler boosting, which seems not to
be the case in gamma-ray binaries. 

Alternatively, relativistic particles can be injected in the surrounding medium
by the wind from a young pulsar \cite{maraschi81}. In this scenario the slowing
rotation of a young pulsar provides stable energy to the non-thermal
relativistic particles in the shocked pulsar wind material outflowing from the
binary companion (see Fig.~\ref{fig:scenario}, right panel). As in one of the
microquasar-jet models, the gamma-ray emission can be produced by inverse
Compton scattering of the relativistic particles from the pulsar wind on
stellar photons. In this context, LS~I~+61~303 would resemble the gamma-ray
binary PSR~B1259$-$63, a radio pulsar in an eccentric orbit around a star of
spectral type Be \cite{aharonian05b}. In fact, recent observations at radio
wavelengths have come in support of the idea that LS~I~+61~303 is a gamma-ray
pulsar rather than a microquasar \cite{dhawan06}. As expected from the pulsar
wind model (see Fig.~\ref{fig:scenario}, right panel), VLBA images of the radio
emission show a relativistic wind from the compact object that spins as a
function of the orbital phase.

The compact objects in these three gamma-ray binaries (LS~5039, LS~I~+61~303
and PSR~B1259$-$63) have eccentric orbits around stars with masses in the
range of 10 to 23 solar masses, and these stars can provide the seed photons
to be scattered by the inverse Compton effect to VHEs. PSR~B1259$-$63 contains
a pulsating neutron star, and for LS~5039 and LS~I~+61~303 the precise mass of
the compact stars is not known. Certainly, they are no more than 5 solar
masses, which is consistent with neutron stars and/or black holes of low mass.
The question that remains open is whether the relativistic particles in
LS~5039 come from accretion powered jets or from the rotational energy of a
pulsar that is spinning down as in PSR~B1259$-$63.

The pulsar wind model requires gamma-ray binaries with neutron stars young
enough to provide large spin down energies. In fact, as in PSR~B1259$-$63,
LS~5039 contains a young compact object. Kinematic studies show that LS~5039
has been shot out from the plane of the Galaxy \cite{ribo02} sometime less than
one million year ago by a supernova explosion produced when the compact object
was formed. Furthermore, it had been proposed that LS~I~+61~303 is a
pulsar-wind source because the time variability and the radio and X-ray spectra
resemble those of young pulsars \cite{dubus06}. Besides, LS~I~+61~303 contains
a Be star like PSR~B1259$-$63, and the high energy emission in both objects
seems to be produced at specific phases of orbital motions of the compact
objects around the Be stars. All Be/X-ray binaries known so far contain neutron
stars and none is known to host a black hole.

However, the jets in LS~5039 seem to be steady and two sided, with bulk motions
of 0.2 to 0.3 times the speed of light, as do the compact jets in black hole
microquasars. Furthermore, in LS~5039 no major radio outbursts are observed
similar to those in PSR~B1259$-$63.

The detection of pulsations would be a definitive proof for the pulsar-wind
mechanism in gamma-ray binaries. On the other hand, detection of VHE emission
from a black hole binary (e.g. Cygnus~X-1, V4641~Sgr, GX~339$-$4) would provide
definitive observational ground to the microquasar-jet model. As done in
\cite{dhawan06} a direct way to distinguish between accretion and rotational
powered gamma-ray binaries is to use radio images with high sensitivity and
angular resolution that would establish clearly whether the high-energy
particles that trigger the VHE emission emanate as pulsar winds or as highly
collimated microquasar jets. 

Gamma-ray binaries are becoming subjects of topical interest in high energy
astrophysics, and their study has important implications. As microquasars, they
would serve as valuable nearby laboratories to gain insight into the physics of
AGN jets. As pulsar-wind gamma-ray binaries they are important because they are
the likely precursors of a much larger population of high-mass X-ray binaries
in the Milky Way. In particular, they may provide clues to understand the early
evolution of the enshrouded hard X-ray binaries being discovered from space
with satellite telescopes such as the European Space Agency's International
Gamma Ray Astrophysics Laboratory ({\it INTEGRAL}).

\end{document}